\documentclass[preprint,aps,nofootinbib]{revtex4}
\usepackage{graphicx}
\usepackage{epstopdf}
\usepackage{subfigure}
\usepackage{textcomp}
\def\bkR{{\rm I\kern-.17em R}}
\def\bkC{{\rm \kern.24em \vrule width.05em height1.4ex depth-.05ex \kern-.26em C}}

\def\be{\beta}

\def\frac#1#2{{\textstyle{{#1}\over {#2}}}}

\def\lsim{\mathrel{\rlap{\lower4pt\hbox{\hskip1pt$\sim$}}
    \raise1pt\hbox{$<$}}}
\def\gsim{\mathrel{\rlap{\lower4pt\hbox{\hskip1pt$\sim$}}
    \raise1pt\hbox{$>$}}}
\def\sqr#1#2{{\vcenter{\vbox{\hrule height.#2pt
         \hbox{\vrule width.#2pt height#1pt \kern#1pt
         \vrule width.#2pt}
         \hrule height.#2pt}}}}

\def\laq{\raise 0.4 ex \hbox{$<$}\kern -0.8 em\lower 0.62 ex\hbox{$\sim$}}
\def\gaq{\raise 0.4 ex \hbox{$>$}\kern -0.7 em\lower 0.62 ex\hbox{$\sim$}}

\def\be{\begin{equation}}
\def\ee{\end{equation}}
\def\beqa{\begin{eqnarray}}
\def\eeqa{\end{eqnarray}}

\def\dalemb#1#2{{\vbox{\hrule height.#2pt
        \hbox{\vrule width.#2pt height#1pt \kern#1pt \vrule width.#2pt}
        \hrule height.#2pt}}}

\def\dalemb#1#2{{\vbox{\hrule height.#2pt
        \hbox{\vrule width.#2pt height#1pt \kern#1pt \vrule width.#2pt}
        \hrule height.#2pt}}}

\def\gtorder{\mathrel{\raise.3ex\hbox{$>$}\mkern-14mu
             \lower0.6ex\hbox{$\sim$}}}
\def\ltorder{\mathrel{\raise.3ex\hbox{$<$}\mkern-14mu
             \lower0.6ex\hbox{$\sim$}}}

\begin{document}

\rightline{August 2012}

\title{Non-linearities in the quantum multiverse}

\author{\vspace*{1cm}Orfeu Bertolami\footnote{Also at Instituto de Plasmas e Fus\~ao Nuclear,
Instituto Superior T\'ecnico, Avenida Rovisco Pais 1, 1049-001 Lisboa, Portugal. E-mail: orfeu.bertolami@fc.up.pt} and Victor Herdeiro\footnote{E-mail: victor.oriedreh@gmail.com}\vspace*{0.5cm}}

\vskip 0.3cm

\affiliation{Departamento de F\'\i sica e Astronomia, Faculdade de Ci\^encias da Universidade do Porto \\
Rua do Campo Alegre 687, 4169-007 Porto, Portugal\vspace*{3.5cm}}


\begin{abstract}

\vskip 0.3cm

{It has been recently proposed that the multiverse of eternal inflation and the many-worlds interpretation of quantum mechanics can be identified, yielding a new view on the measure and measurement problems. In the present note, we argue that a non-linear evolution of observables in the quantum multiverse would be an obstacle for such a description and that these non-linearities are expected from quite general arguments.}

\end{abstract}

\maketitle

\section*{One multiverse to rule them all}

The idea of a multiverse has been repeatedly discussed over the past fifty years in various contexts in physics. Everett's new picture of the measurement problem \cite{Ev}, the so-called Many World Interpretation (MWI) of quantum mechanics, was possibly the first time that this radical concept appeared. The MWI states that upon a measurement performed by an observer on a quantum system, the wave function, splits into all the possible outputs, corresponding each one to an independent universe. In fact, this interpretation pairs up quite well with the scheme of decoherence: the interaction of the quantum system with its environment selects the base of the observables (environment induced selection), such that, after decoherence, the density matrix would describe a superposition of classical states without selecting any of these \textit{outputs}; however, thanks to the MWI, these are all realized in one branch of the split wave function.

Later, in the early 1980's, a development on the then proposed inflationary models, was to consider that in an eternally expanding de-Sitter space where the inflaton was in a metastable state, the ensued first order phase transition would lead to the nucleation of bubbles of the "true" vacua, which could be regarded as a universe with its own Hubble horizon \cite{Linde}. However, on its own, this framework lacks predictability, as an event that will happen in an future eternally expanding multiverse will happen an infinite amount of times, and as such, it is unclear how to compute finite non-arbitrary, i.e. regulated, independent probabilities. This flaw is usually referred to as the measure problem of eternal inflation.

More recently, the multiverse reappeared in the context of string theory, when it was realized that the theory was actually a continuum of theories in the supermoduli-space, the string ``landscape''. The number of possible vacua in this vast space is breathtaking, $G = 10^{100}$ or $10^G$. However, as in our universe the observed value of cosmological constant is $10^{120}$ smaller than its natural value, $M_p^{4}$, $M_p$ being the Planck mass, there must exist a huge number of universes with broken supersymmetry and all possible values for the cosmological constant \cite{BoussP, SussL, WeinM}. Of course, this scenario poses many intriguing questions: How is the vacuum of our world selected? Through anthropic arguments \cite{PolL}? Through quantum cosmology arguments \cite{Holman}? Is the string landscape scenario compatible with predictability \cite{Ellis}? Do the universes of the multiverse interact \cite{Bert2008} (see also Ref. \cite{Linde1988})? Does the multiverse exhibit collective behavior \cite{Alonso}?

A quite interesting recent development is the possibility that these distinct ideas about the multiverse are actually the same \cite{Bousso,Nom}. The main point behind this unification is to attach to an observer a local description of space-time, an idea which has its roots in the so-called black-hole complementarity \cite{Suss}. Thus, in order to describe the nucleation of bubble-universe, the suggestion is to follow the geodesic of an arbitrary observer and to define, within its accessible causal diamond, the decoherence process \cite{Zurek}. This is performed such that the degrees of freedom that escape the future boundary of the causal diamond are inaccessible and are traced over. This leads to the branching of the wave function. Hence, the MWI is reconciled with an eternally inflating space-time.

This can be made a bit more concrete through the generalization of the Born rule \cite{Nom}, for a given observable, ${\cal{O}}_{A_{\mathrm{pre}}}$, before measurement:
\be P(B|A) = \frac{\int\int_{t_1,t_2=0}^{t_1,t_2=+\infty} dt_1dt_2\langle\psi(0)|U(0,t_1){\cal{O}}_{A_{\mathrm{pre}}}U(t_1,t_2){\cal{O}}_{A_{post}\cap B}U(t_2,t_1){\cal{O}}_{A_{pre}}U(t_1,0)|\psi(0)\rangle}{\int\int_{t_1,t_2=0}^{t_1,t_2=+\infty} dt_1dt_2\langle\psi(0)|U(0,t_1){\cal{O}}_{A_{\mathrm{pre}}}U(t_1,t_2){\cal{O}}_{A_{post}}U(t_2,t_1){\cal{O}}_{A_{pre}}U(t_1,0)|\psi(0)\rangle} \ee
where $|\Psi(t)\rangle$ is the wavefunction of our multiverse, starting at $t = 0$ in an inflating state, and $U(t_1,t_2)$ is a unitary evolution operator. It is assumed that every prediction in quantum mechanics should take the form of specific configuration measurements $\{A_{\mathrm{pre}}(t_1),A_{\mathrm{post}}(t_2)\}$, ($t_1$,$t_2$) being natural regulators, and an expected output $B$; the observables, actually projectors operators $\cal{O_{\mathrm{X}}}$, are defined on the full Hilbert space of the multiverse:
\be
{\cal{H}}=\bigoplus_{\cal{M}} {\cal{H_{\cal{M}}}}\hspace*{1cm} \mathrm{where} \hspace*{0.5cm}{\cal{H_{\cal{M}}}}={\cal{H_{\cal{M,\mathrm{bulk}}}}}\otimes{\cal{H_{\cal{M,\mathrm{horizon}}}}}\,,
\ee
where the sum is taken over 4-dimensional geometries, and, is inspired on the Holographic Principle \cite{SussH,Hoof} according to which the dynamics of the degrees of freedom in the bulk and on the horizon are relevant for a consistent description. Hence, this formulation would apply for both instances, the nucleation of a bubble in the context of a specific cosmology or a low-energy quantum mechanical experiment in a universe that has already cooled down. The description of a local observer, necessary to reproduce of the inherent redundancies of general relativity, yields the growth of entanglement between the observed system and its environment, and hence splitting.

Notice that in what concerns the definition of the Hilbert space, one expects that the, observer dependent, separability between the bulk and the horizon subspaces would be blurred in a model where large and small physical scales are mixed due to non-linear evolution. For instance, when defining a quantum state on the bulk, it would be function of the quantum states on the boundary. Of course, the presence of non-linearities would prevent the unitary evolution to realize the Born rule, Eq. (1).

\section*{Non-linear quantum mechanics in the multiverse}

One tacit assumption of the unifying scheme discussed above is the absence of non-linearities in the evolution of the observables and states. The issue of non-linear corrections to quantum mechanics has been recurrently discussed, and, given its generality we will focus on the formalism discussed by Weinberg \cite{Wein,Wein2}. This generalization corresponds to consider extra terms in the Hamiltonian wich are functions $h(\psi_i,\psi^*_k)$ of degree one in the wave-function $\psi_i$ and its conjugate. The study of these non-linear systems has shown some quite interesting departures from quantum mechanics. These include EPR violations, faster-than-light, i.e. Lorentz violating, or backward-in-time signaling, unless one constrains the evolution of a subsystem to depend on a partial tracing over other sub-parts \cite{Pol}. Even more intriguing is that parallel branches of the wave-function would be causally linked, though the so-called Everett phone \cite{Pol}. In the MWI, it would mean that the physics on a universe would depend on the events on every other parallel universes. Furthermore, in an unified MWI/eternally expanding multiverse, our universe would be continuously influenced by every nucleating bubble-universe, rendering the proposed unification, based on the separation Eq. (2) unsuitable to make unambiguous predictions.

Stringent experimental bounds have been set on the presence of these non-linear observables in quantum mechanics \cite{Wein2}. Nonetheless, as pointed out in Ref. \cite{Pol}, some non-linearities, however small, lead a completely different phenomena, rendering the very procedure of separability and performing measurements questionable. Furthermore, a complete argument to rule out this more general dynamics would require to extend the current analysis to high energies.

A possible way to probe this domain is quantum cosmology. The starting point is the Wheeler-DeWitt equation, which yields, when applied to small subsystems, Schr\"odinger equation in the lowest order through a suitable WKB approach in the minisuperspace approximation \cite{Vil}. It has been argued that at the next order in $16 \pi M_p^{-2}$, a correction to Schr\"odinger's equation is found, that can be linearized if the relevant dynamical fields do not depend on the quantum state of the system \cite{Bert}, a rather particular case.

A somewhat more fundamental approach is based on third quantization arguments inspired in string field theory \cite{Banks}. The main ingredient brought by this approach is that the trilinear string field terms render impossible a complete separation between the low-energy and the high-energy degrees of freedom and that this ``mixture'' leads to a Wheeler-DeWitt equation which contains, on quite general grounds, non-linear terms. The general structure of these non-linearities are of the type discussed in Refs. \cite{Wein,Wein2}, which lead to EPR-like violations  \cite{Pol}.

Another relevant point made in Ref. \cite{Banks} is that the interplay between low-energy and high-energy degrees of freedom is precisely what seems to be required to solve the cosmological constant problem (see also Refs. \cite{Cole,Kle}). Therefore, one is lead to conclude that on somewhat general grounds it is rather plausible that at the most fundamental level non-linearities are present in the evolution of the fundamental degrees of freedom describing the dynamics of an observer in space-time. If so, the EPR violations discussed in Ref. \cite{Pol} for non-linear observables are not only a challenge to the MWI, but they turn the unification proposed in Ref. \cite{Bousso} unviable too. This means that the idea that it suffices describing the dynamics of an observer to apprehend the physics of a single universe is untenable. In fact, this is precisely what one should expect if the universes in the multiverse interact among themselves \cite{Linde1988,Bert2008}, which opens up a wide range of new phenomena, including collective behaviour \cite{Alonso}.

\section*{Discussion and Conclusions}

Assuming that a quantum observer ``carries" its space-time, it is possible, through the MWI, to achieve a quantum description of the multiverse that is compatible with a scenario like eternal inflation, as well as, a suitable regularization of the probabilities in the context of this cosmological set up. However, this picture can only be consistent if the branches of the MWI do not interact or communicate with each other. Furthermore, even though, non-linear terms in the Schr\"odinger equation seem to be irrelevant at low energies \cite{Wein2} (see also Ref. \cite{Bert2005}), non-linear terms are expected at higher energies from quantum cosmological or string field theory inspired arguments. In any case, it is likely that the mixture of low-energy and high-energy states seem to be a necessary ingredient to solve the cosmological constant problem, and thus this feature of high-energy models, might lead to non-linear extensions of quantum mechanics. Of course, one cannot disregard the radical possibility that the universes in the multiverse are actually interacting \cite{Linde1988,Bert2008,Alonso}, a conceptual new feature that would turn the effects that we are trying to rule out an inherent property of the multiverse.  

\begin{acknowledgements}

\noindent
The work of OB is partially supported by the Funda\c{c}\~{a}o para a Ci\^{e}ncia e a Tecnologia (FCT) project PTDC/FIS/111362/2009. 

\end{acknowledgements}

\end{document}